  \documentclass[twoside,reqno,11pt]{fcaa-var} %
\usepackage{graphicx}
\usepackage{epsfig}

\usepackage{amsthm}
\usepackage{amsmath}
\usepackage{latexsym}
\usepackage{amsfonts}
\usepackage{amssymb}

\newtheoremstyle{theorem}% name
  {15pt}          % space above
  {15pt}  % space below
  {\sl}  % bofy font
  {\parindent}
  {\sc}  % thm head font
  {. }   % punctuation after thm head
  { }    % space after thm head: `` ``=normal \newline=linebreak
  {}     % thm head specification
\theoremstyle{theorem}

\newtheorem{theorem}{Theorem}[section]

\newtheoremstyle{defi}% name
  {15pt}          % space above
  {15pt}  % space below
  {\rm}  % bofy font
  {\parindent}     % ident - empty=no indent,  \parindent= paragraph indent
  {\sc}  % thm head font
  {. }    % punctuation after thm head
  { }    % space after thm head: `` ``=normal \newline=linebreak
  {}     % thm head specification
\theoremstyle{defi}
\newtheorem{definition}{Definition}[section]

 \def\proofname{\indent\rm P\ r\ o\ o\ f.}
 \def\proofend{\hfill$\Box$}
 \def\qed{\hfill$\Box$} % instead of \square
   \usepackage{hyperref}
   \def\theequation{\arabic{section}.\arabic{equation}}

   \usepackage{graphicx}
\usepackage{amssymb}
\usepackage{dcolumn}% Align table columns on decimal point
\usepackage{bm}% bold math
\newcommand{\be}{\begin{equation}}
\newcommand{\ee}{\end{equation}} 

\allowdisplaybreaks[1]
 \allowdisplaybreaks

 \def\themycopyrightfootnote{\vspace*{3pt}
 \copyright \, Year\,  Diogenes  Co., Sofia
   \par  \noindent pp. xxx--xxx, DOI: ......................
   \hfill  \vspace*{-36pt}
   }
\title[ASYMPTOTIC CYCLES IN FRACTIONAL MAPS \dots ]
      {ASYMPTOTIC CYCLES IN FRACTIONAL MAPS \\ [3pt] OF ARBITRARY POSITIVE ORDERS}
 \author[\normalsize M. Edelman, A.B. Helman]
         {\normalsize Mark Edelman $^1$, Avigayil B. Helman $^2$}

\begin{document}

 \vbox to 1.0cm { \vfill }

 \bigskip %%%\medskip

%%%%%%%%%%%%%%%%%%%%%%%%%%%%%%%%%%%%%
\begin{abstract}

Many natural (biological, physical, etc.) and social systems possess power-law memory, and their mathematical modeling requires application of discrete and continuous fractional calculus. Most of these systems are nonlinear and demonstrate regular and chaotic behavior, which may be very different from the behavior of memoryless systems. Finding periodic solutions is essential for understanding regular and chaotic behavior of nonlinear systems. Fractional systems do not have periodic solutions except fixed points. Instead, they have asymptotically periodic solutions which, in the case of stable regular behavior, converge to the periodic sinks (similar to regular dissipative systems) and, in the case of unstable/chaotic behavior, act as repellers. In one of his recent papers, the first author derived equations which allow calculations of asymptotically periodic points for a wide class of discrete maps with memory. All fractional and fractional difference maps of the orders $0<\alpha<1$ belong to this class.
In this paper we derive the equations that allow calculations of
the coordinates of the asymptotically periodic points for a wider class of maps which include fractional and fractional difference maps of the arbitrary positive orders $\alpha>0$. The maps are defined as convolutions of a generating function $-G_K(x)$, which may be the same as in a corresponding regular map $x_{n+1}=-G_K(x_n)+x_n$,  with a kernel $U_\alpha(k)$, which defines the type of a map. In the case of fractional maps, it is $U_\alpha(k)=k^\alpha$, and it is $U_\alpha(k)=k^{(\alpha)}$, the falling factorial function, in the case of fractional difference maps. In this paper we define the space of kernel functions which allow calculations of the periodic points of the corresponding maps with memory. We also prove that in fractional maps of the orders $1<\alpha<2$ the total of all physical momenta of period--$l$ points is zero.
\medskip


 \smallskip


 \end{abstract} %%%%%%%%%%%%%%%%%%%%

 \maketitle

%%%%%%% end make title %%%%%%%%%%%%%%%%%%%%%%%%%%%%%%%%%%
 \vspace*{-24pt}

\section{Introduction} \label{sec:1} %%%%%% Sect 1 %%%%%%

\setcounter{equation}{0}\setcounter{theorem}{0}

In recent years many authors tried to make fractional generalizations of regular maps. The authors apologize if in the following we are missing some references to outstanding papers on fractional maps. In what follows, the authors intentionally omit references to the papers in which fractional maps are derived as approximate numerical algorithms for solving corresponding fractional differential equations, fractional maps which contain fractions but have no memory (this would contradict the spirit of fractional calculus), and works in which maps are not derived but declared.

\smallskip

Two comprehensive forms of fractional maps, whose introduction has a reasonable justification, are fractional and fractional difference maps described in the review articles \cite{HBV2,HBV4}. Fractional maps were introduced as the exact solutions of fractional differential equations of kicked systems in the same way as the universal map was introduced in regular dynamics by Tarasov and Zaslavsky for systems of the orders $\alpha>1$ \cite{TZ1,TarMap1}. They were extended for orders $0 \le \alpha \le 1$ in \cite{ME1}, investigated in \cite{ME1,ME2,ME3,ME4,ME5,ME6,ME7,ME8,ME9,ME10,ME11}, and reviewed in
\cite{HBV2,HBV4,TarasovBook2011,MER1}. In 2014 Wu, Baleanu, and Zeng proposed to use solutions of fractional difference equations to generate fractional difference maps \cite{Fall,FallC,Peng} and Edelman introduced the universal fractional difference families of maps \cite{ME12,ME13}. Fractional difference maps were investigated in \cite{ME6,ME8,ME9,ME10,ME11,ME12,ME13,AC,Rag,Wang,ME14}.
Applications of fractional and fractional difference maps include economic modeling \cite{TarEc1,TarEc2}, secure communication \cite{Rag}, and modeling of longevity of living species \cite{ME11}.

\smallskip

Fractional and fractional difference maps were first introduced as generalizations of the regular universal and standard maps. These regular maps are closely related to physics and Hamiltonian systems and they are area-preserving. As a result, all integer members of fractional and fractional difference $\alpha$-families of maps with $\alpha>1$ are area/volume preserving and behavior of maps of the order $\alpha$ near $\alpha=2$ is of a special interest. Systems of the order $1<\alpha<2$ behave as systems of the order two but with dissipation. In biology, fractional order $0<\alpha<2$ equations play a significant role in description of electromagnetic waves propagating through the biological tissue (for biological applications of fractional calculus see, e.g., references in review articles \cite{MER1,Bio1,Bio2}). Equations defining periodic points in fractional and fractional difference maps of the orders $0<\alpha<1$ were derived in \cite{ME14}. In this paper we consider equations that allow calculation of asymptotically periodic points in fractional and fractional difference maps of any real orders $\alpha>0$.

\newpage %%%%%%%%%%%%%%%%%%%

\section{Preliminaries} \label{sec:2} %%%%%%% Sect 2 %%%%%%%%%%%%%%

\setcounter{equation}{0}\setcounter{theorem}{0} %%% by VK %%

\subsection{Definition of fractional and fractional difference maps}
\label{sec:2.1}

The definitions (derivations) of fractional and fractional difference families of maps can be found in \cite{HBV2,HBV4} and the references therein.

The fractional Caputo universal map of the order $\alpha$ is defined (derived) as
\vskip -10pt 
% { % \setlength\arraycolsep{0.5pt}
\begin{eqnarray}
x_{n+1}= \sum^{m-1}_{k=0}\frac{b_k}{k!}h^k(n+1)^{k}
-\frac{h^{\alpha}}{\Gamma(\alpha)}\sum^{n}_{k=0} G_K(x_k) (n-k+1)^{\alpha-1},
\label{FrCMapx}
\end{eqnarray} %}
where $\alpha \in \mathbb{R}$, $\alpha \ge 0$, $m=\lceil \alpha \rceil$,  $x_n=x(t=nh)$, $t$ is time, $n \in \mathbb{Z}$, $n \ge 0$ (the same as  $n \in \mathbb{N}_0$), $b_k \in \mathbb{R}$ are constants, and $G_K(x)$ is a function (could be nonlinear) depending on a parameter $K$.

The $h$-difference Caputo universal $\alpha$-family of maps is defined (derived) as
 % {%% \setlength\arraycolsep{0.5pt}
\begin{eqnarray}
x_{n+1} \!=\!
\sum^{m-1}_{k=0}\frac{c_k}{k!}((n\!+\!1)h)^{(k)}_h
-\frac{h^{\alpha}}{\Gamma(\alpha)}
\sum^{n+1-m}_{s=0}\!
(n\!-\!s\!-\!m \!+\! \alpha)^{(\alpha-1)}
G_K(x_{s+m-1}),
\label{FalFacMap_h}
\end{eqnarray} % }
where $x_k=x(kh)$, $\alpha \in \mathbb{R}$, $\alpha \ge 0$, $m=\lceil \alpha \rceil$, $n \in \mathbb{Z}$, $n \ge m-1$, $c_k \in \mathbb{R}$ are constants, and $G_K(x)$ is a function with a parameter $K$. In both maps
$h \in \mathbb{R}$ and $h>0$.
The definition of the falling factorial $t^{(\alpha)}$ is
\begin{equation}
t^{(\alpha)} =\frac{\Gamma(t+1)}{\Gamma(t+1-\alpha)}, \ \ t\ne -1, -2, -3, ...
\label{FrFac}
\end{equation}
The falling factorial is asymptotically a power function:
\begin{equation}
\lim_{t \rightarrow
  \infty}\frac{\Gamma(t+1)}{\Gamma(t+1-\alpha)t^{\alpha}}=1,
\ \ \ \alpha \in  \mathbb{R}.
\label{GammaLimit}
\end{equation}
The $h$-falling factorial $t^{(\alpha)}_h$ is defined as
\begin{equation}
t^{(\alpha)}_h =h^{\alpha}\frac{\Gamma(\frac{t}{h}+1)}{\Gamma(\frac{t}{h}+1-\alpha)}= h^{\alpha}\Bigl(\frac{t}{h}\Bigr)^{(\alpha)}, \ \ \frac{t}{h} \ne -1, -2, -3, ...
\label{hFrFac}
\end{equation}

The important particular forms of the Caputo universal map are the Caputo logistic families of maps with $G_K(x)= x-Kx(1-x)$ which converge to the classical logistic map when $\alpha=1$ and $h=1$
and the Caputo standard families of maps with $G_K(x)= K \sin(x)$, which converge to the classical standard maps when $\alpha=2$ and $h=1$.

\subsection{Asymptotically periodic cycles for $0<\alpha<1$} \label{sec:2.2} %%%%%%%%%%%

Finding periodic trajectories is essential for understanding maps' regular and chaotic dynamics \cite{Cvitanovic}.

It is known that fractional systems may not have periodic solutions except fixed points (see, e.g., \cite{PerD1,PerD2,PerC1,PerC2,PerC3,PerC4,PerC5}) but they may have asymptotically periodic solutions (see \cite{HBV2,HBV4}). A formula for calculation of the asymptotically period two solutions for fractional and fractional difference maps, without a strict proof, may be found in \cite{ME9}. Formulae to find asymptotically periodic points for $0<\alpha<1$ were strictly derived in \cite{ME14}.

When $0<\alpha<1$, all forms of the universal $\alpha$-families of maps
considered in this paper, Eqs.~(\ref{FrCMapx})~and~(\ref{FalFacMap_h}),
can be written as
\begin{eqnarray}
x_{n}= x_0
-\sum^{n-1}_{k=0} {G}^0(x_k) U_{\alpha}(n-k),
\label{FrUUMap}
\end{eqnarray}
where ${G}^0(x)=h^\alpha G_K(x)/\Gamma(\alpha)$ and $x_0$ is the initial condition.
In fractional maps Eq.~(\ref{FrCMapx})
\begin{eqnarray}
U_{\alpha}(n)=n^{\alpha-1}, \ \ \ \  U_{\alpha}(1)=1.
\label{UnFr}
\end{eqnarray}
and in fractional difference maps, Eq.~(\ref{FalFacMap_h}),
{ \setlength\arraycolsep{0.5pt}
\begin{eqnarray}
&&U_{\alpha}(n)=(n+\alpha-2)^{(\alpha-1)}
, \ \ \ \  \nonumber \\
&&U_{\alpha}(1)=(\alpha-1)^{(\alpha-1)}=\Gamma(\alpha).
\label{UnFrDif}
\end{eqnarray}
}
The equations that define asymptotically period $l$ points ($l$-cycles)
\begin{equation}
x_{lim,m}=\lim_{N \rightarrow
  \infty} x_{Nl+m}, \  \  \ 0<m<l+1,
\label{TlpointEx}
\end{equation}
are
\vskip -12pt
{\setlength\arraycolsep{0.5pt}
\begin{eqnarray}
&&x_{lim,m+1}-x_{lim,m}=S_1{G}^0(x_{lim,m})+\sum^{m-1}_{j=1}S_{j+1}{G}^0(x_{lim,m-j})
\nonumber \\
&&+\sum^{l-1}_{j=m}S_{j+1}G^0(x_{lim,m-j+l}), \  \ 0<m<l
\label{LimDifferences}
\end{eqnarray}
}
and
\vskip -10pt
\begin{equation}
\sum^{l}_{j=1}G^0(x_{lim,j})=0,
\label{Close}
\end{equation}
where
\vskip -10pt
\begin{equation}
{S}_{1}=-U_{\alpha}(1) + \sum^{\infty}_{k=1}\Bigl[
U_{\alpha} (lk) - U_{\alpha} (lk+1)\Bigr]
\label{S1}
\end{equation}
and
\vskip -10pt
{\setlength\arraycolsep{0.5pt}
\begin{eqnarray}
&&S_{j+1}=\sum^{\infty}_{k=0}\Bigl[
U_{\alpha} (lk+j) - U_{\alpha} (lk+j+1)\Bigr], \  \ 0<j<l.
\label{Ser}
\end{eqnarray}
}
Note that
\vskip-10pt
\begin{equation}
\sum^{l}_{j=1}S_j=0.
\label{Ssum}
\end{equation}

The results of \cite{ME14} were obtained for a wide class of kernels $U_{\alpha}(x)$ which belong to the space $\mathbb{D}^0(\mathbb{N}_1)$ of diverging functions, defined by the following:

\begin{definition}\label{Def3} %Text of Definition~\ref{Def3}.
A function $f(n)$, $n\in\mathbb{N}_1$, belongs to the space $\mathbb{D}^i(\mathbb{N}_1)$ if $|\sum^{\infty}_{k=1}\Delta^if(k)|>N$, $\forall N$, $N \in \mathbb{N}$, while the series
$\sum^{\infty}_{k=1}\Delta^{i+1}f(k)$ is converging absolutely:
{\setlength\arraycolsep{0.5pt}
\begin{eqnarray}
&&\mathbb{D}^i(\mathbb{N}_1)\ \ = \ \ \{f: |\sum^{\infty}_{k=1}\Delta^if(k)|>N, \ \ \forall N, \ \ N \in \mathbb{N}, \ \
\nonumber \\
&&\sum^{\infty}_{k=1}|\Delta^{i+1}f(k)|=C, \ \ C \in \mathbb{R}_+\}.
\label{DefForm}
\end{eqnarray}
}
\end{definition} %%%%%%%%%%%

Here $\Delta$ is a forward difference operator defined as
\begin{equation}
\Delta f(n)= f(n+1)-f(n).
\label{Delta}
\end{equation}

 We may state % that %
 the following theorem that %
 was proven in \cite{ME14}:

\begin{theorem}\label{Th1} %Text of Theorem~\ref{Th1} ....
If the map Eq.~(\ref{FrUUMap}) with
$U_{\alpha}(n) \in \mathbb{D}^0(\mathbb{N}_1)$ has asymptotically period $l$ points Eq.~(\ref{TlpointEx}) in the domain of the function $G^0(x)$, then they are defined by the system of equations
Eq.~(\ref{LimDifferences}) and Eq.(\ref{Close}), with the coefficients
defined by Eq.~(\ref{S1}) and Eq.(\ref{Ser}).
\end{theorem}

\proof %%%%%%%%%%%%%
The proof is given in \cite{ME14}.
\proofend %%%%%%%%%%

\section{Asymptotically periodic cycles for $1<\alpha<2$} \label{sec:3}

\setcounter{equation}{0}\setcounter{theorem}{0} %% by VK %%

We dedicate a section to the case of fractional order systems with $1<\alpha<2$ which frequently appear in applications. The fractional standard and logistic maps of the orders $1<\alpha<2$ were already investigated in \cite{HBV4,ME1,ME2,ME3,ME4,ME5,ME6,ME9,MER1,Fall,ME12,ME13}. Let's notice that all phase portraits in \cite{Fall} are incorrect.
When $1<\alpha<2$, the map equations, Eq.~(\ref{FrCMapx}) and Eq.(\ref{FalFacMap_h}),
can be written as
\vskip -10pt % { %\setlength\arraycolsep{0.5pt}
\begin{eqnarray}
x_{n+1}= x_0 +h(n+1)p_0
-h\sum^{n}_{k=0} G^1(x_k) U_{\alpha}(n-k+1)+hf_1(n).
\label{FrUUMap2x}
\end{eqnarray} % }
Here $G^1(x)=h^{\alpha-1} G_K(x)/\Gamma(\alpha)$, $x_0$ is the initial coordinate, $U_{\alpha}(n)$
is defined by
Eqs.~(\ref{UnFr})~and~(\ref{UnFrDif}), $p_0$ is the initial
momentum ($b_1$ or $c_1$ in corresponding formulae),
and $f_1(n)=0$
in Eq.~(\ref{FrCMapx}) and
$f_1(n)=h^{\alpha-1} G(x_0)(n-1+\alpha)^{(\alpha-1)}/\Gamma(\alpha) \sim n^{\alpha-1}$ in
Eq.~(\ref{FalFacMap_h}).

\subsection{Equations defining periodic points} \label{sec:3.1} %%%

We define the difference momentum as
\begin{eqnarray}
p^1_{n+1}= \Delta_hx_n=\frac{x_{n+1}-x_{n}}{h}.
\label{p}
\end{eqnarray}
Taking into account that $U_{\alpha}(0)=0$ and using $x_n$ from Eq.~(\ref{FrUUMap2x}), the difference momentum Eq.~(\ref{p})  can be written as
{\setlength\arraycolsep{0.5pt}
\begin{eqnarray}
&&p^1_{n}=p_0
-\sum^{n-1}_{k=0}   G^1(x_k) U^1_{\alpha}(n-k)
+f_1(n-1)-f_1(n-2),
\label{FrUUMap2p}
\end{eqnarray}
}
where
{\setlength\arraycolsep{0.5pt}
\begin{eqnarray}
&&U^1_{\alpha}(n)=U_{\alpha}(n)-U_{\alpha}(n-1 )\nonumber
\\
&&=
\begin{array}{c}
\left\{
\begin{array}{lll}
n^{\alpha-1}-(n-1)^{\alpha-1} \sim n^{\alpha-2}
%\\
{\rm and} \ \ {U}^1_{\alpha}(1)=1
\  \
{\rm in \ \ Eq.~(\ref{FrCMapx})} ;
\\
(n+\alpha -2)^{(\alpha -1)}-(n+\alpha -3)^{(\alpha -1)}
\\
=(\alpha-1)(n+\alpha -3)^{(\alpha-2)} \\
=(\alpha-1)U_{\alpha-1}(n)
\sim n^{\alpha-2}
\\
{\rm and} \ \
{U}^1_{\alpha}(1)=\Gamma(\alpha)
\  \  {\rm in \ \
  Eq.~(\ref{FalFacMap_h})},
\end{array}
\right.
\end{array}
\label{Utilde}
\end{eqnarray}
}
$f_1(n)-f_1(n-1)=0$
in Eq.~(\ref{FrCMapx}) and
$f_1(n)- f_1(n-1) \sim n^{\alpha-2}$ in
Eq.~(\ref{FalFacMap_h}).
Note, that the definitions of ${U}^1_{\alpha}(1)$ in Eq.~(\ref{Utilde}) and
$U_{\alpha}(1)$ in Eqs.~(\ref{UnFr}),~(\ref{UnFrDif}) are identical.

Assuming that asymptotically period $l$ points ($l$-cycles)
Eq.~(\ref{TlpointEx}) exist, the corresponding limiting values of difference momentum are defined as
\begin{equation}
p^1_{lim,m}=\lim_{N \rightarrow
  \infty} p^1_{Nl+m}=\frac{x_{lim,m}-x_{lim,m-1}}{h}, \  \ 1<m<l+1
\label{TlpointEp}
\end{equation}
and
\vskip -10pt
\begin{equation}
p^1_{lim,1}=\lim_{N \rightarrow
  \infty} p^1_{Nl+1}=\frac{x_{lim,1}-x_{lim,l}}{h}.
\label{TlpointEp1}
\end{equation}

The equations defining asymptotically cyclic points are obtained considering the limit $N \rightarrow \infty$, in which case the last two terms of Eq.~(\ref{FrUUMap2p}) disappear. The asymptotic behavior of ${U}^1_{\alpha}(n)$ when $1<\alpha<2$ is the same as the asymptotic behavior of ${U}_{\alpha}(n)$ when $0<\alpha<1$, which means that
$U^1_{\alpha}(x) \in \mathbb{D}^0(\mathbb{N}_1) $. This implies that all steps of the derivation of formulae Eqs.~(\ref{LimDifferences})~-~(\ref{Ssum}), strictly proven in \cite{ME14}, can be repeated here. Subtracting consecutive limiting values and adding all limiting values of difference momenta produces the following results:
{\setlength\arraycolsep{0.5pt}
\begin{eqnarray}
&&p^1_{lim,m+1}-p^1_{lim,m}={S}^1_1{G}^1(x_{lim,m})+\sum^{m-1}_{j=1}{S}^1_{j+1}{G}^1(x_{lim,m-j})
\nonumber \\
&&+\sum^{l-1}_{j=m}{S}^1_{j+1}{G}^1(x_{lim,m-j+l}), \  \ 0<m<l
\label{LimDifferencesP}
\end{eqnarray}
}
and
\vskip -12pt
\begin{equation}
\sum^{l}_{j=1}{G}^1(x_{lim,j})=0,
\label{CloseP}
\end{equation}
where
\begin{equation}
{S}^1_{1}=-{U}^1_{\alpha}(1) + \sum^{\infty}_{k=1}\Bigl[
{U}^1_{\alpha} (lk) - {U}^1_{\alpha} (lk+1)\Bigr],
\label{S1p}
\end{equation}
{\setlength\arraycolsep{0.5pt}
\begin{eqnarray}
&&{S}^1_{j+1}=\sum^{\infty}_{k=0}\Bigl[
{U}^1_{\alpha} (lk+j) - {U}^1_{\alpha} (lk+j+1)\Bigr], \  \ 0<j<l,
\label{SerP}
\end{eqnarray}
}
and, still,
\vskip -12pt
 \begin{equation}
\sum^{l}_{j=1}{S}^1_j=0.
\label{SsumP}
\end{equation}
Using Eqs.~(\ref{TlpointEp})~and~(\ref{TlpointEp1}) the system of $l$ equations which defines all $l$ periodic points can be written as (notice that ${G}^0(x)=h{G}^1(x)$)
{\setlength\arraycolsep{0.5pt}
\begin{eqnarray}
x_{lim,2}-2x_{lim,1}+x_{lim,l} ={S} +\! \sum^{l-1}_{j=1}{S}^1_{j+1}{G}^0(x_{lim,2-j}),
\label{LimDifferencesP1f}
\end{eqnarray}
\vspace*{-5pt} %%%
}
{\setlength\arraycolsep{0.5pt}
\begin{eqnarray}
&&x_{lim,m+1}-2x_{lim,m}+x_{lim,m-1} = {S}^1_1{G}^0(x_{lim,m})
\nonumber \\
&& \hspace*{-0.2cm} %%
+\sum^{m-1}_{j=1}{S}^1_{j+1}{G}^0(x_{lim,m-j})+ \! \sum^{l-1}_{j=m}{S}^1_{j+1}{G}^0(x_{lim,m-j+l}),
\  1<m<l
\label{LimDifferencesPf}
\end{eqnarray}
}
and
\vskip -12pt
\begin{equation}
\sum^{l}_{j=1}{G}^0(x_{lim,j})=0.
\label{ClosePf}
\end{equation}
This is the end of the proof of the following theorem:

\begin{theorem}\label{Th2} %% Th.3.1 %%
If the map Eq.~(\ref{FrUUMap2x}) with $ \Delta U_{\alpha}(n) \in \mathbb{D}^0 (\mathbb{N}_1)$ has asymptotically period $l$ points Eq.~(\ref{TlpointEx}) in the domain of the function $G^1(x)$, then they are defined by the system of equations
Eq.~(\ref{LimDifferencesP1f}), Eq.~(\ref{LimDifferencesPf}),  and Eq.(\ref{ClosePf}), with the coefficients
defined by Eq.~(\ref{S1p}) and Eq.(\ref{SerP}).
\end{theorem}

\subsection{Periodic points in the phase space of discrete fractional systems} \label{sec:3.2} %%%%

Let us notice that in fractional difference maps (truly discrete systems),  values of difference momentum coincide with the corresponding values of physical momentum (see Eqs.~(28)~and~(29) in \cite{ME12}). In this case, after the $x$-coordinates of asymptotically periodic points are found, the limiting values of physical momentum $p$ can be calculated using Eqs.~(\ref{TlpointEp})~and~(\ref{TlpointEp1}). This fact allows us to identify
asymptotically periodic points, $(x_{lim,j},p_{lim,j})$,
for $1 \le j \le l$ in the phase space of a system.

This is not the case in fractional maps, in which a value of physical momentum $p(t)$ is a continuous solution of a differential equation of a kicked system, and $p_n=p(t_n)$, where $t_n$ is a time instant at which a kick occurs. In this case, difference momentum is just an instrument to calculate space coordinates of a cyclic point and a value of physical momentum, which is a solution of the corresponding differential equation, is given by the formula (see Eq.~(18.196) in \cite{TarasovBook2011} and Eq.~(38) in \cite{HBV2})
\vskip -10pt
\begin{equation}
p_{n+1}= p_0-\frac{h^{\alpha-1}}{\Gamma(\alpha-1)}\sum^{n}_{k=0} G_K(x_k) (n-k+1)^{\alpha-2},
\label{FrCMapp2dim}
\end{equation}
\vskip -3pt \noindent 
or
\vskip -12pt
%% {\setlength\arraycolsep{0.5pt}
\begin{eqnarray}
p_{n}= p_0-\sum^{n-1}_{k=0} \tilde{G}^1(x_k) (n-k)^{\tilde{\alpha}-1}
=p_0-\sum^{n-1}_{k=0} \tilde{G}^1(x_k) U_{\tilde{\alpha}} (n-k),
\label{FrCMapp2dimN}
\end{eqnarray} % }
where $\tilde{G}^1(x)=h^{\tilde{\alpha}}G_K(x)/\Gamma(\tilde{\alpha})$ and $\tilde{\alpha}=\alpha-1$. Comparing Eq.~(\ref{FrCMapp2dimN}) to Eq.~(\ref{FrUUMap}), we may derive equations similar to Eq.~(\ref{LimDifferences}) that define relationships between limiting values of momentum
\begin{equation}
p_{lim,m}=\lim_{N \rightarrow
  \infty} p_{Nl+m}, \  \  \ 0<m<l+1,
\label{TlpointEpn}
\end{equation}
in asymptotically period $l$ cycles:
\vskip -12pt
{\setlength\arraycolsep{0.5pt}
\begin{eqnarray}
&&p_{lim,m+1}-p_{lim,m}=S_{1,{\tilde{\alpha}}}\tilde{G}^1 (x_{lim,m})+\sum^{m-1}_{j=1}S_{j+1,{\tilde{\alpha}}}\tilde{G}^1 (x_{lim,m-j})
\nonumber \\
&&+\sum^{l-1}_{j=m}S_{j+1,{\tilde{\alpha}}}\tilde{G}^1 (x_{lim,m-j+l}), \  \ 0<m<l,
\label{LimDifferencesPp}
\end{eqnarray}
}
where $S_{j,{\tilde{\alpha}}}$ is the value of $S_j$ calculated for $\alpha=\tilde{\alpha}$. These equations define only relative values of
momentum $p_{lim,m}$ of $l$-cycles. The equation that closes the system is a consequence of the following theorem:

\begin{theorem}\label{Th3} %% Th. 3.2 %%%%%%%%
In fractional maps of the orders $1<\alpha<2$ the total of all physical momenta of period--$l$ points is zero:
\begin{equation}
\sum^{l}_{j=1}p_{lim,j}=0, \  \ l \in \mathbb{Z}, \  \ l>1.
\label{ClosePfn}
\end{equation}
\end{theorem}

\proof %%%%%%%%%%%%%

Let us notice that from the definition Eq.~(\ref{p}) follows that
in a period $l$ point the total of all difference momenta is zero:
\begin{equation}
\sum^{l}_{j=1}p^1_{lim,j}=0, \   \ l \in \mathbb{Z}, \  \ l>1.
\label{PDtotal}
\end{equation}
Using equations Eqs.~(\ref{FrUUMap2p})~and~(\ref{FrCMapp2dimN}) and ignoring $f_1(n-1)-f_1(n-2)$ terms which disappear in the limit $n \rightarrow \infty$,  we may write that
\begin{equation}
p_n-p^1_n=\sum^{n}_{m=1}\tilde{G}^1(x_{n-m})  \Bigl[ \frac{m^{\alpha-1}- (m-1)^{\alpha-1}}{\alpha-1}- m^{\alpha-2} \Bigr].
\label{pminusp1}
\end{equation}
Now let us consider a total of $l$ consecutive differences at large values of $n$:
$$ 
\lim_{N \rightarrow \infty} \sum^{Nl+l}_{n=Nl+1}(p_n-p^1_n) \hspace*{7cm} %%%
$$ %%%% by VK %%%
\vspace*{-6pt} 
\begin{equation}
= \lim_{N \rightarrow \infty}\sum^{Nl+l}_{n=Nl+1}
\sum^{n}_{m=1}\tilde{G}^1(x_{n-m})  \Bigl[ \frac{m^{\alpha-1}- (m-1)^{\alpha-1}}{\alpha-1}- m^{\alpha-2} \Bigr].
\label{pminusp1l}
\end{equation}
After changing the order of summation on the right hand side (RHS), this equality can be written as
%% {\setlength\arraycolsep{0.5pt}
\begin{eqnarray}
&& \sum^{l}_{j=1}p_{lim,j}-\sum^{l}_{j=1}p^1_{lim,j}=\lim_{N \rightarrow \infty} \sum^{Nl+l}_{m=1}\Bigl[ \frac{m^{\alpha-1}- (m-1)^{\alpha-1}}{\alpha-1}- m^{\alpha-2} \Bigr] 
 %% \sum^{Nl+l}_{n=Nl+1}\tilde{G}^1(x_{n-m}) %% 
\nonumber \\
&&
\times \sum^{Nl+l}_{n=Nl+1}\tilde{G}^1(x_{n-m}) %% VK %% 
+\lim_{N \rightarrow \infty} \sum^{Nl+l}_{m=Nl+1}\Bigl[ \frac{m^{\alpha-1}- (m-1)^{\alpha-1}}{\alpha-1}- m^{\alpha-2} \Bigr] 
\nonumber \\
&& \times  \sum^{Nl+l}_{n=m}\tilde{G}^1(x_{n-m}).
\label{pminusp1ln}
\end{eqnarray} %% }
The expressions in the square brackets on the RHS of this equation are of the order $m^{\alpha-3}$. The very last sum of $\tilde{G}^1(x_{n-m})$ in the equation contains not more than $l$ bounded terms and is bounded by some constant $C_1$. Its multiplier in the square brackets is of the order $m^{\alpha-3}$ and for any $\varepsilon$ there exists $N_{l1}$ such that
for any $N>N_{l1}$ this multiplier will be less than $\varepsilon/(3 lC_1)$. Then, the whole last term will be less than $\varepsilon/3$. The first term on the RHS may be split into two terms:
{\setlength\arraycolsep{0.5pt}
\begin{eqnarray}
&&  \lim_{N \rightarrow \infty} \sum^{Nl+l}_{m=1}\Bigl[ \frac{m^{\alpha-1}- (m-1)^{\alpha-1}}{\alpha-1}- m^{\alpha-2} \Bigr] \sum^{Nl+l}_{n=Nl+1}\tilde{G}^1(x_{n-m})  =
\nonumber \\
&&\lim_{N_1\rightarrow \infty,N-N_1\rightarrow \infty } \sum^{N_1l+l}_{m=1}\Bigl[\frac{2-\alpha}{2}m^{\alpha-3}+O(m^{\alpha-4}) \Bigr] \sum^{Nl+l}_{n=Nl+1}\tilde{G}^1(x_{n-m})
%\nonumber
\\
&&+\lim_{N_1\rightarrow \infty,N-N_1\rightarrow \infty } \sum^{Nl+l}_{m= N_1l+l+1 }\Bigl[\frac{2-\alpha}{2}m^{\alpha-3}+O(m^{\alpha-4}) \Bigr] \sum^{Nl+l}_{n=Nl+1}\tilde{G}^1(x_{n-m}),
\nonumber
\label{pminusp1part}
\end{eqnarray}
}
where $N>N_1$.
The sum of $l$ consecutive values of $\tilde{G}^1(x_{n-m})$ is bounded by some constant $C_2$ (the limit of this  sum is zero, see Eq.~(\ref{ClosePf})). The series of the terms in the square brackets is converging absolutely. So, the series itself is bounded by some constant $C_3$ and for any $\varepsilon$ there exists $N_{l2}$ such that for any $N_1>N_{l2}$ the total of the terms of this series starting from $N_1l+l+1$ will be less than $\varepsilon/(3C_2)$.  Then, the magnitude of the last term (the third line of the equation) can be made less than $\varepsilon/3$ by selecting a large enough value of $N_1$. For any $\varepsilon$ we may find a large $N_{l3}$ such that, for any $N>N_{l3}$, $N-N_1$ will be large enough to make the sum of $l$ consecutive values of $\tilde{G}^1(x_{n-m})$ less than $\varepsilon/(3C_3)$. Then, the first term on the RHS of  Eq.~(\ref{pminusp1part}) (the second line of the equation) will be less than $\varepsilon/3$.

Now, for any $\varepsilon$, if we choose a large $N$ ($N>N_{l1}$ and $N>N_{l2}$), then the expression on the RHS of Eq.~(\ref{pminusp1ln})) will be less than $\varepsilon$, which means that the expression on the left is zero and
\vskip -10pt
\begin{equation}
\sum^{l}_{j=1}p_{lim,j}=\sum^{l}_{j=1}p^1_{lim,j}=0, \   \ l \in \mathbb{Z}, \  \ l>1.
\label{ClosePfnn}
\end{equation}
This ends the proof.
\proofend %%%%%%%%%%

The results of Theorem~\ref{Th2} were previously observed in numerical calculations, starting from \cite{ME2} (see also review \cite{MER1}), but has never been proven before.

As in the case $\alpha<1$, to calculate periodic points, one should first calculate the sums ${S}^1_j$.

\subsection{Calculation of the sums ${S}^1_j$} \label{sec:3.3} %%%%%%%%%%%%%%%%

Calculation of the sums ${S}^1_j$, in the case of fractional difference maps, is quite simple. Let us notice that in this case, the combination of equations
Eqs.~(\ref{S1p}),~(\ref{SerP}),~(\ref{Utilde}),~(\ref{S1}),~and~(\ref{Ser}) produces the following results:
{\setlength\arraycolsep{0.5pt}
$$ 
{S}^1_1(\alpha)=-\Gamma(\alpha)+(2-\alpha)(\alpha-1)\sum^{\infty}_{k=1}\frac{\Gamma(lk+\alpha-2)}{\Gamma(lk+1)}
\hspace*{1cm} %%% VK %%
$$ 
\begin{eqnarray}
%% {S}^1_1(\alpha)=-\Gamma(\alpha)+(2-\alpha)(\alpha-1)\sum^{\infty}_{k=1}\frac{\Gamma(lk+\alpha-2)}{\Gamma(lk+1)} %
 =(\alpha-1)S_1(\alpha-1) \hspace*{5cm} %%% ?? %%
\label{St1vsS1FD}
\end{eqnarray}
}
and
\vskip -12pt 
{\setlength\arraycolsep{0.5pt}
\begin{eqnarray}
&&{S}^1_{j+1}(\alpha)=(2-\alpha)(\alpha-
1)\sum^{\infty}_{k=0}\frac{\Gamma(lk+j+\alpha-2)}{\Gamma(lk+j+1)}
\nonumber \\
&& 
=(\alpha-1)S_{j+1}(\alpha-1), \  \  \  0<j<l.
\label{StvsSFD}
\end{eqnarray}
}
In the last two formulae we explicitly show the dependence of sums on $\alpha$ by adding it as the argument. Because $0<\alpha-1<1$, it is possible to use numerical formulae and tables derived in \cite{ME14} to obtain $S_j$ and then ${S}^1_j$.

We will follow the steps described in \cite{ME14} to calculate the sums in the case of fractional maps. In this case, Eqs.~(\ref{S1p})~and (\ref{SerP}) can be written as
\vskip -12pt 
{\setlength\arraycolsep{0.5pt}
\begin{eqnarray}
&&{S}^1_{1}=-1 + \sum^{N}_{k=1}\Bigl[
2(lk)^{\alpha-1}-(lk-1)^{\alpha-1}- (lk+1)^{\alpha-1} \Bigr]
\nonumber \\
&&+ \sum^{\infty}_{k=N+1}\Bigl[
2(lk)^{\alpha-1}-(lk-1)^{\alpha-1}- (lk+1)^{\alpha-1} \Bigr]
\label{S1pn}
\end{eqnarray}
}
and
\vskip -10pt %% {\setlength\arraycolsep{0.5pt}
$$ %%% \begin{eqnarray}
%% &&
{S}^1_{j+1}=\sum^{N}_{k=0}\Bigl[
2(lk)^{\alpha+j-1} - (lk+j-1)^{\alpha-1} - (lk+j+1)^{\alpha-1} \Bigr] 
$$ %% \nonumber \\ && 
\vspace*{-7pt}
\begin{equation} %%%
+\! \sum^{\infty}_{k=N+1}\! \Bigl[
2(lk)^{\alpha+j-1} \!-\!(lk+j-1)^{\alpha-1}
\!-\! (lk+j+1)^{\alpha-1} \Bigr],\  0<j<l,
\label{SerPn}
\end{equation} %%%% \end{eqnarray}%% }
where $N$ is a large number (we used $N=20000$ in all calculations).
In the second (the infinite) sums we factor out $(lk)^{\alpha-1}$ and develop the difference into a Taylor series. After some simple transformations, the expression to calculate ${S}^1_{j}$ can be written as
{\setlength\arraycolsep{0.5pt}
\begin{eqnarray}
&&{S}^1_{1}=-1 + \sum^{N}_{k=1}\Bigl[
2(lk)^{\alpha-1}-(lk-1)^{\alpha-1}- (lk+1)^{\alpha-1} \Bigr]
\nonumber \\
&&+(1-\alpha)(\alpha-2)l^{\alpha-3}\Biggl[\zeta_N(3-\alpha)
%\\ &&
+\frac{(\alpha-3)(\alpha-4)}{12l^2}\zeta_N(5-\alpha) \Biggr] \\
&&+ \, O(N^{\alpha-6})
\nonumber
\label{S1pnf}
\end{eqnarray}
}
and
\vskip -10 pt %% {\setlength\arraycolsep{0.5pt}
\begin{eqnarray}
&&{S}^1_{j+1}=\sum^{N}_{k=0}\Bigl[
2(lk+j)^{\alpha-1}-(lk+j-1)^{\alpha-1} - (lk+j+1)^{\alpha-1} \Bigr] \nonumber \\
&&+(1-\alpha)(\alpha-2)l^{\alpha-3}\Biggl\{\zeta_N(3-\alpha) +\frac{\alpha-3}{l}\Biggl[j\zeta_N(4-\alpha)
\\
&& +\frac{\alpha-4}{12l}\Biggl((6j^2+1)\zeta_N(5-\alpha)
+\frac{(\alpha-5)j(2j^2+1)}{l}
\zeta_N(6-\alpha)\Biggr)\Biggr]\Biggr\} \nonumber \\
&& +\, O(N^{\alpha-6}),\nonumber
\label{SerPnf}
\end{eqnarray}%% }
where
\vskip -13pt
\begin{equation}
\zeta_N(m-\alpha)=\zeta(m-\alpha)-\sum^{N}_{k=1}k^{\alpha-m},
\label{ZetaNl}
\end{equation}
$0<j<l$, and we used a fast method for calculating values of the Riemann $\zeta$-function. To estimate the accuracy of our computations, we also calculated the value of $\Sigma S^1_i$, whose deviation from zero represents the absolute error. Researchers, who apply and investigate fractional maps may use the results of the $S^1_j$ constants calculations for the period two and three points presented in
Tables~\ref{table:T2_Fr}~and~\ref{table:T3_FrMT1}. It may be noticed that, in the period two case, the definition of the constant $S^1_2$ coincides with the definition of the introduced in \cite{ME2} constant $V_{\alpha l}$.

\begin{table}[ht!] %%%%%%%%%% Table 1 %%%%%%%%%%%%%%%%%%%%%%%%%%%
\centering
%\resizebox{1.4\textwidth}{!}{\begin{minipage}{\textwidth}
    \begin{tabular}{| c  | c       |      c  | c      |}
    \hline
    $\alpha$ & $S^1_1$ & $S^1_2$ &  $\Sigma S^1_i$    \\ \hline
    1.01  & -.9954781& .9954781& 4.44e-16 \\ \hline
    1.05  & -.9772693& .9772693& 2.89e-15 \\ \hline
    1.1   & -.9542397& .9542397& 6.55e-15 \\ \hline
    1.15  & -.9309180& .9309180& 1.30e-14 \\ \hline
    1.2   & -.9073118& .9073118& 1.67e-14 \\ \hline
    1.25  & -.8834292& .8834292& 2.16e-14 \\ \hline
    1.3   & -.8592791& .8592791& 3.72e-14 \\ \hline
    1.35  & -.8348713& .8348713& 3.01e-14 \\ \hline
    1.4   & -.8102162& .8102162& 3.71e-14 \\ \hline
    1.45  & -.7853250& .7853250& 5.47e-14 \\ \hline
    1.5   & -.7602096& .7602096& 5.62e-14 \\ \hline
    1.55  & -.7348831& .7348831& 6.73e-14 \\ \hline
    1.6   & -.7093590& .7093590& 7.38e-14 \\ \hline
    1.65  & -.6836521& .6836521& -7.55e-15 \\ \hline
    1.7   & -.6577777& .6577777& -2.78e-13 \\ \hline
    1.75  & -.6317523& .6317523& -9.10e-13 \\ \hline
    1.8   & -.6055932& .6055932& -2.04e-13 \\ \hline
    1.85  & -.5793187& .5793187& -3.74e-13 \\ \hline
    1.9   & -.5529480& .5529480& -2.76e-12 \\ \hline
    1.95  & -.5265014& .5265014& -1.05e-11 \\ \hline
    1.99  & -.5053036& .5053036& -6.53e-12 \\ \hline
    \end{tabular}
    \medskip %%VK %%
    \caption{The values of $S^1_j$, $1 \le j \le 2$, for T=2 cycles (fractional maps).}
    \label{table:T2_Fr}
%\end{minipage} }
\end{table} %%%%%%%%%%%%%%%%%%%%%%%%%%%%

\begin{table}[ht!] %%%%% Table 2 %%%%%%%
\centering
%\resizebox{1.4\textwidth}{!}{\begin{minipage}{\textwidth}
    \begin{tabular}{| c  | c       |      c  | c      | c       |}
    \hline
    $\alpha$ & $S^1_1$ & $S^1_2$ & $S^1_3$  &  $\Sigma S^1_i$    \\ \hline
    1.01  & -.998088676 & .994297820 & .003790856 & 5.67E-16\\ \hline
    1.05  & -.990205877 & .971261837 & .018944039 & 2.48E-15\\ \hline
    1.1   & -.979805615 & .941958610 & .037847005 & 5.56E-15\\ \hline
    1.15  & -.968778551 & .912096431 & .056682119 & 1.15E-14\\ \hline
    1.2   & -.957104032 & .881682864 & .075421169 & 1.36E-14\\ \hline
    1.25  & -.944761480 & .850727005 & .094034474 & 2.07E-14\\ \hline
    1.3   & -.931730456 & .819239584 & .112490872 & 2.78E-14\\ \hline
    1.35  & -.917990742 & .787233052 & .130757690 & 2.03E-14\\ \hline
    1.4   & -.903522418 & .754721680 & .148800738 & 3.69E-14\\ \hline
    1.45  & -.888305951 & .721721659 & .166584292 & 9.76E-14\\ \hline
    1.5   & -.872322287 & .688251202 & .184071085 & 3.86E-14\\ \hline
    1.55  & -.855552955 & .654330642 & .201222312 & 1.79E-13\\ \hline
    1.6   & -.837980174 & .619982544 & .217997630 & 2.31E-14\\ \hline
    1.65  & -.819586970 & .585231803 & .234355167 & 1.86E-14\\ \hline
    1.7   & -.800357297 & .550105757 & .250251540 & -4.96E-13\\ \hline
    1.75  & -.780276168 & .514634289 & .265641879 & -8.29E-13\\ \hline
    1.8   & -.759329798 & .478849937 & .280479860 & 2.76E-13\\ \hline
    1.85  & -.737505746 & .442788001 & .294717744 & 2.04E-12\\ \hline
    1.9   & -.714793081 & .406486651 & .308306429 & 6.40E-12\\ \hline
    1.95  & -.691182527 & .369987024 & .321195502 & -4.32E-12\\ \hline
    1.99  & -.671642567 & .340674171 & .330968396 & 8.96E-12\\ \hline
    \end{tabular}
    \medskip %% VK %%
    \caption{The values of $S^1_j$, $1 \le j \le 3$, for period 3 cycles (fractional maps).}
    \label{table:T3_FrMT1}
%\end{minipage} }
\end{table} %%%%%%%%%%%%%%%%%%%%%%%%%%%%%

\section{Asymptotically periodic cycles for $\alpha>2$} \label{sec:4}%%%%%%%%%%%%%%%%%%%%

\setcounter{equation}{0}\setcounter{theorem}{0}

Let us consider an arbitrary $m-1<\alpha<m$. The fractional Caputo universal map Eq.~(\ref{FrCMapx}) can be written as
\begin{equation}
x_{n+1}= \sum^{m-1}_{k=0}\frac{b_k}{k!}h^k(n+1)^{k}
-h^{m-1}\sum^{n}_{k=0} G^{m-1}(x_k) U_{\alpha}(n-k+1),
\label{FrCMapxnn}
\end{equation}
where  $n \in \mathbb{N}_0$, $U_{\alpha}(n)$ is defined by
Eq.~(\ref{UnFr}), and 
\vskip -12pt 
$$G^{m-1}(x)= G_K(x)h^{\alpha-m+1}/\Gamma(\alpha).$$ %% VK %%

The $h$-difference Caputo universal map Eq.~(\ref{FalFacMap_h}) can be written as
{\setlength\arraycolsep{0.5pt}
\begin{eqnarray}
&& x_{n+1}= x_0+\sum^{m-1}_{k=1}\frac{c_k}{k!}h^k(n+1)n...(n+2-k)
\nonumber \\
&& -h^{m-1}
\sum^{n}_{k=0}
 G^{m-1}(x_k) U_{\alpha}(n-k+1)
 \nonumber \\ %% VK %% 
&& + h^{m-1}
\sum^{m-2}_{k=0}
 G^{m-1}(x_k) U_{\alpha}(n-k+1),
\label{FrDifCMapxnn}
\end{eqnarray}
}
where $x_k=x(kh)$, $\alpha \in \mathbb{R}$, $\alpha \ge 0$, $m=\lceil \alpha \rceil$, $n \in \mathbb{N}_{m-1}$, $c_k \in \mathbb{R}$, and $U_{\alpha}(n)$ is defined by
Eq.~(\ref{UnFrDif}).

\subsection{Higher order difference momenta} \label{sec:4.1} %%%%%%%%%%%%%%%%%%%

Let us define order $i>1$ difference momentum as
\begin{eqnarray}
p^i_{n+1}= \frac{p^{i-1}_{n+1}-p^{i-1}_{n}}{h}=\Delta^i x_{n-i+1},  \    \ 0<i<m.
\label{pi}
\end{eqnarray}
To derive the equations that define periodic points we'll need to derive the expression for the limiting ($n \rightarrow \infty$) values of the $m-1$st order momentum $p^{m-1}$. The last term (the sum) on the RHS of
Eq.~(\ref{FrDifCMapxnn}) is of the order $n^{\alpha-1}$ and after the application of the $m-1$st order of the $h$-difference operator it will be of the order $n^{\alpha-m}$. This term will tend to zero in the limit $n \rightarrow \infty$ and may be ignored. Now, let uss consider the $m-1$st order $h$-difference of the first term (which accounts for the initial conditions) on the RHS of Eq.~(\ref{FrCMapxnn}) written for $x_n$:
\begin{equation}
\sum^{m-1}_{k=0}\frac{b_k}{k!}h^{k-m+1}
\sum^{m-1}_{i=0}(-1)^i
\left( \begin{array}{c}
m-1 \\ i
\end{array} \right)
(n-i)^{k}.
\label{FrCMapxnnIC}
\end{equation}
First, let us prove the following theorem (identity):

\begin{theorem}\label{Id1} %% Th. 4.1 %%
For any $n \in \mathbb{Z}$, $m \in \mathbb{N}_0$, and
$k \in \mathbb{N}_0$ (when $k=0$, $n \notin [0,m]$)
the following identity holds:
{\setlength\arraycolsep{0.5pt}
\begin{eqnarray}
&& \sum^{m}_{i=0}(-1)^i
\left( \begin{array}{c}
m \\ i
\end{array} \right)
(n-i)^{k}
=
\begin{array}{c}
\left\{
\begin{array}{lll}
0  \     \
%\\
{\rm for}\  \ 0 \le k <m\ ,
\\
k! \     \
{\rm for}\  \ k =m\ .
\end{array}
\right.
\end{array}
\label{id1}
\end{eqnarray}
}
\end{theorem}

\proof %%%%%%%%%%%%%
We prove the identity by induction. When $m=0$, $m=1$, or $m=2$, for $0 \le k \le m$ the identity is easily verifiable. In what follows we use the Pascal identity
\begin{equation}
\left( \begin{array}{c}
m \\ i
\end{array} \right)
=
\left( \begin{array}{c}
m-1 \\ i-1
\end{array} \right)
+
\left( \begin{array}{c}
m-1 \\ i
\end{array} \right).
\label{CombId}
\end{equation}
Assuming that the identity Eq.~(\ref{id1}) holds for the total with the positive upper limit $m-1$, we may write:
{\setlength\arraycolsep{0.5pt}
\begin{eqnarray}
&& \sum^{m}_{i=0}(-1)^i
\left( \begin{array}{c}
m \\ i
\end{array} \right)
(n-i)^{k}
= n^k+
\sum^{m}_{i=1}(-1)^i
\left( \begin{array}{c}
m \\ i
\end{array} \right)
(n-i)^{k}
\nonumber \\
&&=n^k+
\sum^{m}_{i=1}(-1)^i
\left( \begin{array}{c}
m-1 \\ i-1
\end{array} \right)
(n-i)^{k}+
\sum^{m}_{i=0}(-1)^i
\left( \begin{array}{c}
m-1 \\ i
\end{array} \right)
(n-i)^{k}
-n^k
\nonumber \\
&&= -\!
\sum^{m-1}_{j=0}(-1)^j
\left( \begin{array}{c}
m-1 \\ j
\end{array} \right)
((n\!-\!1)\!-\!j)^{k}\!
+
\sum^{m-1}_{i=0}(-1)^i
\left( \begin{array}{c}
m-1 \\ i
\end{array} \right)
(n\!-\!i)^{k},
\label{id1N}
\end{eqnarray}
}
where $j=i-1$. If $k \le m$ then, according to our assumption, the two totals on the last line in Eq~(\ref{id1N}) are identical and have the opposite signs and the result is zero. If $k=m$, then
{\setlength\arraycolsep{0.5pt}
\begin{eqnarray}
&& \sum^{m}_{i=0}(-1)^i
\left( \begin{array}{c}
m \\ i
\end{array} \right)
(n-i)^{k}
= \sum^{m-1}_{i=0}(-1)^i
\left( \begin{array}{c}
m-1 \\ i
\end{array} \right)
\{(n-i)^{m}-[(n-i)-1]^m\}
\nonumber \\
&&
= \sum^{m-1}_{i=0}(-1)^i
\left( \begin{array}{c}
m-1 \\ i
\end{array} \right)
\Biggl[m(n-i)^{m-1}-\frac{m(m-1)}{2}(n-i)^{m-2}+\ldots\Biggr]
\nonumber \\
&&
= m\sum^{m-1}_{i=0}(-1)^i
\left( \begin{array}{c}
m-1 \\ i
\end{array} \right)
(n-i)^{m-1}
\nonumber \\
&& %%% VK  %%%%
-\frac{m(m-1)}{2}
\sum^{m-1}_{i=0}(-1)^i
\left( \begin{array}{c}
m-1 \\ i
\end{array} \right)
(n-i)^{m-2}+\ldots
\nonumber \\
&&
=m(m-1)!+0+\ldots= m!=k!.
\label{id1NNN}
\end{eqnarray}
}
This ends the proof.
\proofend %%%%%%%%%%

\smallskip

As a result of this theorem, we may conclude that the $m-1$st order $h$-difference Eq.~(\ref{FrCMapxnnIC}) is equal to $b_{m-1}$, which is the initial value of the $m-1$st derivative of $x$.

In the case of the fractional difference map, the the $m-1$st order $h$-difference of the first term in the expression for $x_n$ can be written like Eq.~(\ref{FrCMapxnnIC}), but with the falling factorial instead the power function:
\begin{equation}
\sum^{m-1}_{k=0}\frac{c_k}{k!}h^{k-m+1}
\sum^{m-1}_{i=0}(-1)^i
\left( \begin{array}{c}
m-1 \\ i
\end{array} \right)
(n-i)^{(k)}.
\label{FrDifCMapxnnIC}
\end{equation}
The proof by induction of the following theorem, is similar to the proof of Theorem~\ref{Id1}:

\begin{theorem}\label{Id2} %% Th.4.2 %%%
For any $n \in \mathbb{Z}_+$ and $m \in \mathbb{Z}_+$, $n>m$, and
$k \in \mathbb{Z}_+$ the following identity holds:
{\setlength\arraycolsep{0.5pt}
\begin{eqnarray}
&& \sum^{m}_{i=0}(-1)^i
\left( \begin{array}{c}
m \\ i
\end{array} \right)
(n-i)^{(k)}
=
\begin{array}{c}
\left\{
\begin{array}{lll}
0  \     \
%\\
{\rm for}\  \ 0 < k <m\ ,
\\
k! \     \
{\rm for}\  \ k =m\ .
\end{array}
\right.
\end{array}
\label{id2}
\end{eqnarray}
}
\end{theorem}

\proof %%%%%%%%%%%%%
When $m=2$ and $k$ is 1 or 2 the identity holds.
Let us assume that the theorem holds for the total with the upper limit $m-1$, then:
%% {\setlength\arraycolsep{0.5pt}
$$ %% \begin{eqnarray} && 
\sum^{m}_{i=0}(-1)^i
\left( \begin{array}{c}
m \\ i
\end{array} \right)
(n-i)^{(k)}
= n^{(k)}+
\sum^{m}_{i=1}(-1)^i
\left( \begin{array}{c}
m \\ i
\end{array} \right)
(n-i)^{(k)}
$$ \vspace*{-6pt} %% \nonumber \\ && %% 
$$ 
=n^{(k)}+
\sum^{m}_{i=1}(-1)^i
\left( \begin{array}{c}
m-1 \\ i-1
\end{array} \right)
(n-i)^{(k)}+
\sum^{m}_{i=0}(-1)^i
\left( \begin{array}{c}
m-1 \\ i
\end{array} \right)
(n-i)^{(k)}
-n^{(k)}
$$ \vspace*{-6pt} %% \nonumber \\ &&= %%%%%
\begin{equation} %%%%
-  \sum^{m-1}_{j=0}(-1)^j
\left( \begin{array}{c}
m-1 \\ j
\end{array} \right)
((n\!-\!1)\!-\!j)^{(k)}+ \!
\sum^{m-1}_{i=0}(-1)^i
\left( \begin{array}{c}
m-1 \\ i
\end{array} \right)
(n\!-\!i)^{(k)},
\label{id2N}
\end{equation} %%% \end{eqnarray} %% }
where $j=i-1$. If $k \le m$ then, according to our assumption, the two totals on the last line in Eq~(\ref{id2N}) are identical and have the opposite signs and the result is zero. If $k=m$, then
 %% {\setlength\arraycolsep{0.5pt}
$$ %%% \begin{eqnarray} && 
\sum^{m}_{i=0}(-1)^i
\left( \begin{array}{c}
m \\ i
\end{array} \right)
(n-i)^{(k)}
= \sum^{m-1}_{i=0}(-1)^i
\left( \begin{array}{c}
m-1 \\ i
\end{array} \right)
\{(n-i)^{(m)}\!-\![(n-i)\!-\!1]^{(m)}\}
$$ \vspace*{-6pt} %% \nonumber \\ && %%%
\begin{equation} %% (4.12) 
= m\sum^{m-1}_{i=0}(-1)^i
\left( \begin{array}{c}
m-1 \\ i
\end{array} \right)
(n-1-i)_{(m-1)}
=m(m-1)!= m!=k!.
\label{id2NNN}
\end{equation} %%% \end{eqnarray} %% }
This ends the proof.
\proofend %%%%%%%%%%

We see that the $m-1$st order $h$-difference Eq.~(\ref{FrDifCMapxnnIC}) is equal to $c_{m-1}$. Accumulating all the results discussed in this section, in both fractional and fractional difference cases, the expression for $p^{m-1}_n$ can be written as
{\setlength\arraycolsep{0.5pt}
\begin{eqnarray}
&&p^{m-1}_{n}=p_{m-1,0}
-\sum^{n-1}_{k=0}   G^{m-1}(x_k) U^{m-1}_{\alpha}(n-k)
+F(n),
\label{FrUUMapMT2p}
\end{eqnarray}
}
where $p_{m-1,0}$ is a constant, $F(n) \rightarrow 0$ when
$n \rightarrow \infty$ and the definition of $U^{m-1}_{\alpha}$ is
{\setlength\arraycolsep{0.5pt}
\begin{eqnarray}
&&U^{m-1}_{\alpha}(n)=U^{m-2}_{\alpha}(n)-U^{m-2}_{\alpha}(n-1 )
=\Delta^{m-1} U_{\alpha}(n-m+1).
\label{UtildeM}
\end{eqnarray}
}
In the case of fractional difference maps, the equality
\begin{equation}
U_{\alpha}(n)=0, \  \ {\rm when} \ \ n<1
\label{UNeg}
\end{equation}
follows from the definition Eq.~(\ref{UnFrDif}). In fractional maps with $\alpha>1$ this is true only for $n=0$ and we will postulate this equality for $n<0$.

In fractional maps,
{\setlength\arraycolsep{0.5pt}
\begin{eqnarray}
&& U^{m-1}_{\alpha}(n)=
\begin{array}{c}
\left\{
\begin{array}{lll}
\sum^{m-1}_{i=0}(-1)^i
\left( \begin{array}{c}
m-1 \\ i
\end{array} \right)
(n-i)^{\alpha-1} \   \
{\rm for} \ \ n > m-1,
\\
\sum^{n-1}_{i=0}(-1)^i
\left( \begin{array}{c}
m-1 \\ i
\end{array} \right)
(n-i)^{\alpha-1} \   \
{\rm for} \  \ n < m.
\end{array}
\right.
\end{array}
\label{UtildeMFr}
\end{eqnarray}
}

In fractional difference maps, assuming that the identity
\begin{equation}
U^{m-2}_{\alpha}(n)
=(\alpha-1)^{(m-2)}U_{\alpha-m+2}(n),
\label{UtildeMFrDifAs}
\end{equation}
which, according to Eq.~(\ref{Utilde}), holds for $m=3$, is true, we may prove by induction that this identity holds for any $m$:
{\setlength\arraycolsep{0.5pt}
\begin{eqnarray}
&&U^{m-1}_{\alpha}(n)= U^{m-2}_{\alpha}(n)-U^{m-2}_{\alpha}(n-1)
\nonumber \\ %% VK %% 
&& =
(\alpha-1)^{(m-2)}[U_{\alpha-m+2}(n)-U_{\alpha-m+2}(n-1)]
\nonumber \\
&& =
(\alpha-1)^{(m-2)}U^1_{\alpha-m+2}(n)
=
(\alpha-1)^{(m-1)}U_{\alpha-m+1}(n)
\nonumber \\ %% VK %% 
&& = \frac{\Gamma(\alpha) \Gamma(n+\alpha-m)}{\Gamma(\alpha-m+1) \Gamma(n)}.
\label{UtildeMFrDif}
\end{eqnarray}
}
For large $n$, $U^{m-1}_{\alpha}(n)\sim n^{\alpha-m}$ in both, fractional and fractional difference cases.

\subsection{Equations defining periodic points}\label{sec:4.2} %%%%%%%%%%%%%

Assuming that asymptotically period $l$ points ($l$-cycles)
Eq.~(\ref{TlpointEx}) exist, the corresponding limiting values of the order $i<m=\lceil \alpha \rceil$ difference momentum are defined as
\begin{equation}
p^i_{lim,k}=\lim_{N \rightarrow
  \infty} p^i_{Nl+k}=\frac{p^{i-1}_{lim,k}-p^{i-1}_{lim,k-1}}{h}, \  \ 1<k<l+1
\label{TlpointEpHO}
\end{equation}
and
\vskip -12pt
\begin{equation}
p^i_{lim,1}=\lim_{N \rightarrow
  \infty} p^{i}_{Nl+1}=\frac{p^{i-1}_{lim,1}-p^{i-1}_{lim,l}}{h},
\label{TlpointEp1HO}
\end{equation}
which can be written as
{\setlength\arraycolsep{0.5pt}
\begin{eqnarray}
&& p^i_{lim,k}=
\begin{array}{c}
\left\{
\begin{array}{lll}
h^{-i}\sum^{i}_{j=0}(-1)^j
\left( \begin{array}{c}
i \\ j
\end{array} \right)
x_{lim,k-j} \   \   \  \
{\rm for} \ \ i<k<l+1;
\\
 h^{-i}\sum^{k-1}_{j=0}(-1)^j
\left( \begin{array}{c}
i \\ j
\end{array} \right)
x_{lim,k-j}
\\
+h^{-i}\sum^{i}_{j=k}(-1)^j
\left( \begin{array}{c}
i \\ j
\end{array} \right)
x_{lim,l-j+k} \   \
{\rm for} \ \ 0<k \le i.
\end{array}
\right.
\end{array}
\label{TlpointEpHOAll}
\end{eqnarray}
}
The equations defining asymptotically cyclic points are obtained considering the limit $N \rightarrow \infty$, in which case the last term of Eq.~(\ref{FrUUMapMT2p}) disappears.
If we assume that $U_{\alpha}(n) \in \mathbb{D}^{m-1}(\mathbb{N}_1)$ for $m-1<\alpha<m$, then from the definition of $U^{m-1}_{\alpha}(n)$ Eq.~(\ref{UtildeM}) (with the extension Eq.~(\ref{UNeg})) and
Definition~\ref{Def3} follows that $U^{m-1}_{\alpha}(x) \in \mathbb{D}^0(\mathbb{N}_1)$.
This implies that all steps of the derivation of formulae Eqs.~(\ref{LimDifferences})~-~(\ref{Ssum}), strictly proven in \cite{ME14}, can be repeated here. Subtracting consecutive limiting values and adding all limiting values of the $m-1$st difference momenta produces the following results:
{\setlength\arraycolsep{0.5pt}
\begin{eqnarray}
&&p^{m-1}_{lim,k+1}-p^{m-1}_{lim,k}={S}^{m-1}_1{G}^{m-1} (x_{lim,k})+\sum^{k-1}_{j=1}{S}^{m-1}_{j+1}{G}^{m-1} (x_{lim,k-j})
\nonumber \\
&&+\sum^{l-1}_{j=k}{S}^{m-1}_{j+1}{G}^{m-1} (x_{lim,k-j+l}), \  \ 0<k<l
\label{LimDifferencesPm}
\end{eqnarray}
}
and
\vskip -13pt
\begin{equation}
\sum^{l}_{j=1}{G}^{m-1}(x_{lim,j})=0,
\label{ClosePm}
\end{equation}
where
\vskip -12pt
\begin{equation}
{S}^{m-1}_{1}=-{U}^{m-1}_{\alpha}(1) + \sum^{\infty}_{k=1}\Bigl[
{U}^{m-1}_{\alpha} (lk) - {U}^{m-1}_{\alpha} (lk+1)\Bigr],
\label{S1pm}
\end{equation}
\vspace*{-8pt} %%%%
{\setlength\arraycolsep{0.5pt}
\begin{eqnarray}
&&{S}^{m-1}_{j+1}=\sum^{\infty}_{k=0}\Bigl[
{U}^{m-1}_{\alpha} (lk+j) - {U}^{m-1}_{\alpha} (lk+j+1)\Bigr], \  \ 0<j<l,
\label{SerPm}
\end{eqnarray}
}
and, still,
\vskip -12pt
 \begin{equation}
\sum^{l}_{j=1}{S}^{m-1}_j=0.
\label{SsumPn}
\end{equation}
Using Eq.~(\ref{TlpointEpHOAll}) the system of $l$ equations which defines all $l$ periodic points after some transformations and using the Pascal identity can be written as (notice that
${G}^0(x)=h^{m-1}{G}^{m-1}(x)$),
%% {\setlength\arraycolsep{0.5pt}
$$ %% \begin{eqnarray} &&
x_{lim,k+1}+(-1)^{m}x_{lim,k-m+1}-
\sum^{m-2}_{j=0}(-1)^j
\left( \begin{array}{c}
m \\ j+1
\end{array} \right)
x_{lim,k-j}
={S}^{m-1}_1{G}^{0} (x_{lim,k})
$$ \vspace*{-6pt}%% \nonumber \\ &&
\begin{equation}%%
+\sum^{k-1}_{j=1}{S}^{m-1}_{j+1}{G}^{0} (x_{lim,k-j})
+\sum^{l-1}_{j=k}{S}^{m-1}_{j+1}{G}^{0} (x_{lim,k-j+l}), \  m-1<k<l,
\label{LimDifferencesPmm1}
\end{equation} %% \end{eqnarray} %% }
%% {\setlength\arraycolsep{0.5pt}
$$ %% \begin{eqnarray} &&
x_{lim,m}+(-1)^{m}x_{lim,l}+
\sum^{m-1}_{j=1}(-1)^j
\left( \begin{array}{c}
m \\ j
\end{array} \right)
x_{lim,m-j}
={S}^{m-1}_1{G}^{0} (x_{lim,m-1})
$$ \vspace*{-6pt} %% \nonumber \\ && %%
\begin{equation} %%
+\sum^{m-2}_{j=1}{S}^{m-1}_{j+1}{G}^{0} (x_{lim,m-1-j})
+ \!
\sum^{l-1}_{j=m-1}{S}^{m-1}_{j+1}{G}^{0} (x_{lim,m-1-j+l}), \  k=m-1,
\label{LimDifferencesPmm2}
\end{equation} %% \end{eqnarray} %% }
%% {\setlength\arraycolsep{0.5pt}
$$ % \begin{eqnarray}&& %%
x_{lim,k+1}+(-1)^{m}x_{lim,l+k-m+1}-
\sum^{k-1}_{j=1}(-1)^j
\left( \begin{array}{c}
m \\ j+1
\end{array} \right)
x_{k-j}
$$ \vspace*{-6pt} %% \nonumber \\ && %%% 
$$
-\sum^{m-2}_{j=k}(-1)^j
\left( \begin{array}{c}
m \\ j+1
\end{array} \right)
x_{lim,l+k-j}
={S}^{m-1}_1{G}^{0} (x_{lim,k})
+\sum^{k-1}_{j=1}{S}^{m-1}_{j+1}{G}^{0} (x_{lim,k-j})
$$ \vspace*{-6pt} %% \nonumber \\ && %%
\begin{equation} %%
+\sum^{l-1}_{j=k}{S}^{m-1}_{j+1}{G}^{0} (x_{lim,k-j+l}), \   0<k<m-1,
\label{LimDifferencesPmm3}
\end{equation} %%% \end{eqnarray} %% }
and
\vskip -12pt 
\begin{equation}
\sum^{l}_{j=1}{G}^0(x_{lim,j})=0.
\label{ClosePfm1}
\end{equation}
This is the end of the proof of the following theorem:

\begin{theorem}\label{Th2m} %% Th. 4.3 %%%
If the map
\begin{equation}
x_{n+1}= f(n)
-h^{m-1}\sum^{n}_{k=0} G^{m-1}(x_k) U_{\alpha}(n-k+1),
\label{FrCMapxm}
\end{equation}
where $\lim_{n \rightarrow \infty} \Delta^{m-1}f(n)$ is a constant, $m=\lceil \alpha \rceil$, $\alpha \in \mathbb{R}$, $\alpha>0$, $h>0$, and \break %%%%
$\Delta^{m-1}U_{\alpha}(n) \in \mathbb{D}^{0}(\mathbb{N}_1)$ has asymptotically period $l$ points Eq.~(\ref{TlpointEx}) in the domain of the function $G^{m-1}(x)$, then they are defined by the system of equations
Eqs.~(\ref{LimDifferencesPmm1}) -- Eq.~(\ref{ClosePfm1}), with the coefficients defined by Eq.~(\ref{S1pm}) and Eq.(\ref{SerPm}).
\end{theorem}

{\sc Remark}.\,
When $m>1$, to use Eq.~(\ref{S1pm}) and Eq.(\ref{SerPm}), the domain $U(n)$ is extended to $n \in \mathbb{N}_{2-m}$ and it is assumed that $U(n)=0$ for $1-m<n<1$.

\medskip

Fractional and fractional difference maps belong to the class of maps to which the theorem applies.

\subsection{Calculation of the sums $S^{m-1}_j$} \label{sec:4.3} %%%%%%%%%%%%%

Calculation of the sums $S^{m-1}_j$, in the case of fractional difference maps, is very simple. In this case, combination of equations
Eq.~(\ref{UtildeMFrDif})~and~Eq.~(\ref{SerPm}) (which with $j=0$ is equivalent to Eq.~(\ref{S1pm})) produces the following result:
{\setlength\arraycolsep{0.5pt}
\begin{eqnarray}
&&S^{m-1}_{j+1}(\alpha)=(m-\alpha)\sum^{\infty}_{k=0}\frac{\Gamma(\alpha)\Gamma(lk+j+\alpha-m)}{ \Gamma(\alpha-m+1)\Gamma(lk+j+1)}
=(\alpha-1)S^{m-2}_{j+1}(\alpha-1)
\nonumber \\
&&=(\alpha-1)^{(m-1)}S_{j+1}(\alpha-m+1)
, \  \  \  0 \le j<l,
\label{St1vsS1FDm}
\end{eqnarray}
}
where we explicitly show the dependence of the sums on $\alpha$ by adding it as the argument. Because $0<\alpha-m+1<1$, it is possible to use numerical formulae and tables derived in \cite{ME14} to obtain $S_j$ and then ${S}^{m-1}_j$.

The calculations are more complicated in the case of fractional maps:
{\setlength\arraycolsep{0.5pt}
\begin{eqnarray}
&&{S}^{m-1}_{1}=-{U}^{m-1}_{\alpha}(1) + \sum^{\infty}_{k=1}\Bigl[
{U}^{m-1}_{\alpha} (lk) - {U}^{m-1}_{\alpha} (lk+1)\Bigr]
\nonumber \\
&&=-1+\sum^{\lceil \frac{m-1}{l} \rceil -1 }_{k=1}\Bigl\{
\sum^{lk-1}_{i=0}(-1)^i
\left( \begin{array}{c}
m-1 \\ i
\end{array} \right)
\Bigl[
(lk-i)^{\alpha-1}-(lk-i+1)^{\alpha-1}
\Bigr]
\nonumber \\ 
&&-(-1)^{lk}
\left( \begin{array}{c}
m-1 \\ lk
\end{array} \right)
\Bigl\}
+\delta_{\lceil \frac{m-1}{l} \rceil [\frac{m-1}{l}]}
\Bigl[ \sum^{lk-1}_{i=0}(-1)^i
\left( \begin{array}{c}
m-1 \\ i
\end{array} \right)
(lk-i)^{\alpha-1} %%%
\nonumber \\
&&
-\sum^{m-1}_{i=0}(-1)^i
\left( \begin{array}{c}
m-1 \\ i
\end{array} \right)
(lk-i+1)^{\alpha-1}
\Bigr]
%\nonumber
\\
&&+\sum^{N}_{k=[\frac{m-1}{l}]+1}
\sum^{m-1}_{i=0}(-1)^i
\left( \begin{array}{c}
m-1 \\ i
\end{array} \right)
\Bigl[
(lk-i)^{\alpha-1}-(lk-i+1)^{\alpha-1}
\Bigr]
\nonumber \\
&&+\sum^{\infty}_{k=N+1}
\sum^{m-1}_{i=0}(-1)^i
\left( \begin{array}{c}
m-1 \\ i
\end{array} \right)
\Bigl[
(lk-i)^{\alpha-1}-(lk-i+1)^{\alpha-1}
\Bigr] \nonumber
\label{S1pmcalc}
\end{eqnarray}
}
and  %%%%%%%%%%%%%%%%%%%%%%%%%%  
{\setlength\arraycolsep{0.5pt}
\begin{eqnarray}
&&{S}^{m-1}_{j+1}=\sum^{\infty}_{k=0}\Bigl[
{U}^{m-1}_{\alpha} (lk+j) - {U}^{m-1}_{\alpha} (lk+j+1)\Bigr]
\nonumber \\
&&=\sum^{\lceil \frac{m-j-1}{l} \rceil -1 }_{k=0}\Bigl\{
\sum^{lk+j-1}_{i=0}(-1)^i
\left( \begin{array}{c}
m-1 \\ i
\end{array} \right)
\Bigl[
(lk+j-i)^{\alpha-1}-(lk+j-i+1)^{\alpha-1}
\Bigr]
\nonumber \\  %%%%%%%%%%%%%%%%%%%%%%%%%%%%%%%%%%%%%%
&&-(-1)^{lk+j}
\left( \begin{array}{c}
m-1 \\ lk+j
\end{array} \right)
\Bigl\}
+\delta_{\lceil \frac{m-j-1}{l} \rceil [\frac{m-j-1}{l}]}
\Bigl[ \sum^{lk+j-1}_{i=0}(-1)^i
\left( \begin{array}{c}
m-1 \\ i
\end{array} \right)
%%  (lk+j-i)^{\alpha-1} %%% 
\nonumber \\
&&
\times\, (lk+j-i)^{\alpha-1} 
-\sum^{m-1}_{i=0}(-1)^i
\left( \begin{array}{c}
m-1 \\ i
\end{array} \right)
(lk+j-i+1)^{\alpha-1}
\Bigr]
\\
&&
+\sum^{N}_{k=[\frac{m-j-1}{l}]+1}
\sum^{m-1}_{i=0}(-1)^i
\left( \begin{array}{c}
m-1 \\ i
\end{array} \right)
\Bigl[
(lk+j-i)^{\alpha-1}-(lk+j-i+1)^{\alpha-1}
\Bigr]
\nonumber \\
&&
+ \!\! \sum^{\infty}_{k=N+1}\! \sum^{m-1}_{i=0}(-1)^i
\left( \begin{array}{c}
m\!-\!1 \\ i
\end{array} \right)\!
\Bigl[
(lk+j-i)^{\alpha-1} \!-\! (lk+j-i+1)^{\alpha-1}
\Bigr], 
 \ 0<j<l, \nonumber
\label{SerPmcalc}
\end{eqnarray}
}
where $\delta_{mn}$ is the Kronecker delta and $N$ is a large number (in all our calculations we assumed $N=20000$). Both expressions,
Eq.~(\ref{S1pmcalc}) and Eq.~(\ref{SerPmcalc}), can be written as the totals of two sums $S^{m-1}_{j+1,fin}$ and $S^{m-1}_{j+1,inf}$ ($0 \le j<l$). $S^{m-1}_{j+1,fin}$ is the sum of a large but finite number of terms (all lines in the corresponding formulae except the last one) and can be computed very fast. To calculate the second sum we use the Taylor expansions, results of Theorem~\ref{Id1}, and changes in the order of summation:
{\setlength\arraycolsep{0.5pt}
\begin{eqnarray}
&&{S}^{m-1}_{j+1,inf}= \!
\sum^{\infty}_{k=N+1}\! \sum^{m-1}_{i=0}(-1)^i
\left( \begin{array}{c}
m-1 \\ i
\end{array} \right)
\Bigl[
(lk+j-i)^{\alpha-1}-(lk+j-i+1)^{\alpha-1}
\Bigr]
\nonumber \\
&&=\sum^{\infty}_{k=N+1}
\sum^{m-1}_{i=0}(-1)^i
\left( \begin{array}{c}
m-1 \\ i
\end{array} \right)
(lk)^{\alpha-1}
\sum^{\infty}_{q=1}
\frac{(\alpha-1)^{(q)}}{q!(lk)^q}
[(j-i)^q-(j-i+1)^q]
\nonumber
\\
&&=-\sum^{\infty}_{k=N+1} \sum^{\infty}_{q=1}
(lk)^{\alpha-1}
\frac{(\alpha-1)^{(q)}}{q!(lk)^q}
\sum^{m-1}_{i=0}(-1)^i
\left( \begin{array}{c}
m-1 \\ i
\end{array} \right)
\sum^{q-1}_{p=0}
\left( \begin{array}{c}
q \\ p
\end{array} \right)
(j-i)^p
\\
&&=-\sum^{\infty}_{q=1}
\frac{(\alpha-1)^{(q)}}{q!}
\sum^{\infty}_{k=N+1}
(lk)^{\alpha-1-q}
\sum^{q-1}_{p=0}
\left( \begin{array}{c}
q \\ p
\end{array} \right)
\sum^{m-1}_{i=0}(-1)^i
\left( \begin{array}{c}
m-1 \\ i
\end{array} \right)
(j-i)^p
\nonumber
\\
&&=-\sum^{\infty}_{q=m}
l^{\alpha-1-q}
\frac{(\alpha-1)^{(q)}}{q!}
\sum^{q-1}_{p=m-1}
\left( \begin{array}{c}
q \\ p
\end{array} \right)
\sum^{m-1}_{i=0}(-1)^i
\left( \begin{array}{c}
m-1 \\ i
\end{array} \right)
(j-i)^p
%% \sum^{\infty}_{k=N+1} k^{\alpha-1-q} %% 
\nonumber
\\
&&
\times\, \sum^{\infty}_{k=N+1} k^{\alpha-1-q} 
= -\sum^{\infty}_{q=m}
l^{\alpha-1-q}
\frac{(\alpha-1)^{(q)}}{q!} \nonumber \\ %%%%
&& \times\, \sum^{q-1}_{p=m-1}
\left( \begin{array}{c}
q \\ p
\end{array} \right)
\sum^{m-1}_{i=0}(-1)^i
\left( \begin{array}{c}
m-1 \\ i
\end{array} \right)
(j-i)^p
\zeta_N(q+1-\alpha),
\nonumber
\label{SerPmcalcInf}
\end{eqnarray}
}
where $0 \le j < l$ and $\zeta_N(x)$ is defined by Eq.~(\ref{ZetaNl}). Finally, %%
we may write
{\setlength\arraycolsep{0.5pt}
\begin{eqnarray}
&&{S}^{m-1}_{j+1}=
{S}^{m-1}_{j+1,fin}
- l^{\alpha-1-m}
(\alpha-1)^{(m)}
\zeta_N(m+1-\alpha)
\nonumber \\
&&-\sum^{m+3}_{q=m+1}
l^{\alpha-1-q}
\frac{(\alpha-1)^{(q)}}{q!}
\sum^{q-1}_{p=m-1}
\left( \begin{array}{c}
q \\ p
\end{array} \right)
\sum^{m-1}_{i=0}(-1)^i
\left( \begin{array}{c}
m-1 \\ i
\end{array} \right)
(j-i)^p \nonumber \\ %%%%
&& \times\, \zeta_N(q+1-\alpha)
%% \nonumber \\ && %%
+ O(N^{\alpha-m-4 }), \  \  \ 0 \le j<l,
\label{SerPmcalcInfFin}
\end{eqnarray}
}
which allows high accuracy calculations of the sums.
It is easy to verify that in the case $m=2$, this expression coincides
with Eq.~(\ref{S1pnf}) and Eq.~(\ref{SerPnf}).

\section{A note on generalized discrete fractional calculus} \label{sec:5}

\setcounter{equation}{0}\setcounter{theorem}{0} %% by VK %%

The notion of the generalized fractional calculus was introduced in \cite{Koch1} and further developed in multiple papers including \cite{Koch2,Koch3,Koch4,Luch1,Luch2,Luch3,Tar1,Tar2}. The general fractional integrals and derivatives were defined using Bernstein functions and Sonine kernels to satisfy the fundamental theorems. In \cite{Tar2} the author proposed the general fractional (universal) maps as the exact solutions of equations with a general fractional derivative and periodic kicks.

Fractional sums and differences and discrete fractional calculus were introduced at the end of the last century \cite{GZ,MR}. In \cite{Fall} the authors proposed to use solutions of fractional difference equations to generate fractional difference maps, which kernels are different from the kernels of fractional maps obtained as solutions of fractional differential equations with periodic kicks. During the last few years various types of generalizations of fractional differential and difference operators were introduced. The authors did not find any references to papers on distributed and variable fractional order generalizations of fractional difference operators, but
we may expect that solutions of fractional difference equations of distributed (with the kernels integrated over $m-1< \alpha <m$) and variable orders ($m-1< \alpha(t) <m$) will be of the form
Eq.~(\ref{FrCMapxm}) and satisfy the conditions of Theorem~\ref{Th2m}.

Kernels $U_{\alpha}(n) \in \mathbb{D}^{m-1}(\mathbb{N}_1)$
were introduced to allow finding asymptotically periodic solutions of
Eq.~(\ref{FrCMapxm}). We may call the map defined by Eq.~(\ref{FrCMapxm}), in which
$\lim_{n \rightarrow \infty} \Delta^{m-1}f(n)$ is a constant, $m=\lceil \alpha \rceil$, $\alpha \in \mathbb{R}$, $\alpha>0$, $h>0$, and $\Delta^{m-1}U_{\alpha}(n) \in \mathbb{D}^{0}(\mathbb{N}_1)$, the general fractional $h$-difference map.
Finding and investigating of the operator $\Delta_{(U),h}^t$, such that the map Eq.~(\ref{FrCMapxm}) is the solution of the fractional $h$-difference equation of the form
\begin{equation}
\Delta_{(U),h}^tx(t)=-G(x(t)),
\label{Gen}
\end{equation}
and the relationship between the kernels which belong to $\mathbb{D}^{m}$ and the kernels of the general fractional calculus is an interesting problem, but it is not the subject of the current paper.

\section{Conclusion} \label{sec:6} %%%%%%%%%%%%%%%%%%%%%%%%%%

Let us summarize the main new results obtained in this paper.

The main goal of this paper was to derive the equations which define asymptotically periodic points in a wide class of maps with memory. The maps are defined as convolutions with kernels $U_{\alpha}(x)$ which belong to the space $\mathbb{D}^{m-1}(\mathbb{Z})$ (or $\Delta^{m-1}U_{\alpha}(x) \in \mathbb{D}^{0}(\mathbb{Z})$) defined by Definition~\ref{Def3}. Fractional and fractional difference maps of real positive orders $\alpha$ belong to this class of maps with $m=\lceil \alpha \rceil$. Theorem~\ref{Th2m} formulates the equation and is the main achievement of the paper. Particular forms of the equations for the cases of $0 < \alpha <1$ and
$1 < \alpha <2$, which have important in applications, are also formulated.

The coefficients of the derived system of equations Eqs.~(\ref{LimDifferencesPmm1})-- Eq.~(\ref{ClosePfm1}) are obtained as slowly converging series  defined by Eq.~(\ref{S1pm}) and Eq.(\ref{SerPm}). The formulae which allow fast calculations of the coefficients in the cases of fractional and fractional difference maps are given by Eq.~(35) from \cite{ME14} and Eq.~(\ref{St1vsS1FDm}) (for fractional difference maps) and by Eq.~(\ref{S1pmcalc}), Eq.~(\ref{SerPmcalc}), and Eq.~(\ref{SerPmcalcInfFin}) (for fractional maps). The versions of these formulae for the case $0 < \alpha <2$ are also obtained and used to calculate the tables of coefficients:  
Tables~1--5 from \cite{ME14} and Tables~\ref{table:T2_Fr}~and~\ref{table:T3_FrMT1}.

The stable periodic points represent the asymptotic behavior of nonlinear systems with memory and, in the case of chaotic systems, the periodic points represent skeletons of chaos \cite{Cvitanovic,May}.
Fractional modifications (addition of the power-law memory) of conservative regular systems behave similarly to the corresponding regular systems with dissipation (see, e.g., \cite{ZSE}). In these cases, periodic points will be dense in the corresponding chaotic attractors.
Our results will be useful to any researcher who works with fractional/fractional difference maps to calculate the periodic points.

Another result, formulated in Theorem~\ref{Th3}, is the proof of the conjecture that in fractional maps of the orders $1<\alpha<2$ the total of all physical momenta of period--$l$ points is zero. This conjecture was previously used in investigation of the fractional standard and logistic maps.

One of the fundamental questions is whether there exists a self-similarity in bifurcation diagrams of nonlinear fractional systems near transition to chaos. Now, when we have the means to numerically find periodic points and define bifurcation points, we may try to answer this question. In the following paper we intend to numerically analyze the bifurcation behavior of the solutions of the derived equations for the case of the fractional and fractional difference logistic maps in order to make a conjecture about the existence or non-existence of the Feigenbaum numbers and their possible values.

\section*{Acknowledgements} %%%%%%%%%%%%%%%%%%%%%%

The author acknowledges support from Yeshiva University's  2021--2022 Faculty Research Fund
and expresses his gratitude to the administration of Courant
Institute of Mathematical Sciences at NYU
for the opportunity to perform some of the computations at Courant.

%\section*{Conflict of interest}
%The authors declare that they have no conflict of interest.

%%%%%%%%%%%%%%%%%%%%%%%%%%%%%%%%%%%%%%%%%%%%%%%%%%%%%%%%%%%%%%%%%%%%%%
 %%%%%%%%%%%%%%%%%%%%%%%%%%%%%%%%%%%%%%%%%%%%%%%%%%%%%%%%

\bigskip % \smallskip

 \it

 \noindent
   %(First) Author's full postal address
$^1$ Department of Physics, %  \\
Stern College at Yeshiva University \\
245 Lexington Ave, New York, NY 10016, USA \\
Courant Institute of Mathematical Sciences \\
New York University \\
251 Mercer St., New York, NY 10012, USA\\
Department of Mathematics\ BCC, CUNY \\
2155 University Avenue, Bronx, New York 10453, USA \\[4pt]
  e-mail: medelma1@yu.edu \,  (Corresponding author)\\ 
  %%%% M. Edelman: https://orcid.org/0000-0002-5190-3651  %%%
\hspace*{0.2cm} \hfill Received: ................. , Revised: November 13, 2021 \\[12pt]%% ??????
  % Second Author's address
$^2$ Stern College at Yeshiva University \\
245 Lexington Ave, New York, NY 10016, USA \\[4pt]
  e-mail: ahelman@mail.yu.edu

\end{document}